\documentclass[11pt]{article}
\pdfoutput =1
\usepackage{graphics}
\usepackage{graphicx}
\usepackage{subcaption}
\DeclareGraphicsExtensions{.pdf}
\usepackage{float} 
\usepackage{subfloat}
\usepackage{amsmath}
\usepackage{slashed}
\usepackage{amsfonts}  
\usepackage{amssymb}
\usepackage{physics}

\def\th{\theta}

\usepackage{comment}
\usepackage{color}

\usepackage{subcaption}
\begin{document}
\date{}
\title{{\bf{\Large Phases of deformed sine dilaton gravity and JT limit
}}}
\author{
{\bf { ~Paramesh Mahapatra}$
$\thanks{E-mail:  paramesh\_m@ph.iitr.ac.in}}\\
 {\normalsize  Department of Physics, Indian Institute of Technology Roorkee,}\\
  {\normalsize Roorkee 247667, Uttarakhand, India}
\\[0.3cm]
{\bf { Hemant Rathi}$ $\thanks{E-mail: hemant.rathi@krmangalam.edu.in}}\\ 
{\normalsize School of Basic and Applied Sciences, K.R. Mangalam University,}\\
{\normalsize Gurugram 122103, Haryana, India}
  \\[0.3cm]
 {\bf { ~Dibakar Roychowdhury}$
$\thanks{E-mail:  dibakar.roychowdhury@ph.iitr.ac.in}}\\
 {\normalsize  Department of Physics, Indian Institute of Technology Roorkee,}\\
  {\normalsize Roorkee 247667, Uttarakhand, India}
\\[0.3cm]
}
\maketitle
\abstract{We study the implications of marginal deformation in sine dilaton gravity (sDG). We investigate the thermodynamic properties of deformed sine dilaton gravity (dsDG) and examine the phase stability of black holes in two space-time dimensions. We found that the deformation parameter ($\lambda$) sets a critical temperature threshold ($T_0$). Below this critical value ($T<T_0$), the system enters an unphysical phase characterized by a fractional density of states and a negative Wald entropy. Above the critical temperature ($T>T_0$), we obtain a physical black hole phase with positive entropy and negative free energy. Furthermore, we evaluate the Komar integral, derive the Casimir function for dsDG, and establish the first law of black hole thermodynamics. Finally, we explore the infrared (IR) limit of dsDG which we identify as Jackiw-Teitelboim (JT) gravity accompanied by marginal deformation. We also perform a detailed analysis of various thermodynamic entities pertinent to this deformed JT solution.} 

\section{Introduction and overview}
In recent years, the two-dimensional gravitational model known as Jackiw-Teitelboim (JT) gravity \cite{Teitelboim:1983ux}-\cite{Maldacena:2016upp} has played a significant role as the simplest paradigm for understanding the gauge/gravity duality \cite{Maldacena:1997re}-\cite{Polchinski:2010hw}, with its boundary physics described by the one-dimensional quantum mechanical Sachdev-Ye-Kitaev (SYK) model \cite{Sachdev:1992fk}-\cite{Maldacena:2016hyu}. JT gravity is a two-dimensional Einstein gravity coupled with a scalar field (known as dilaton) and a negative cosmological constant. As an exactly solvable model of 2D quantum gravity, JT gravity provides valuable insights into quantum gravity and black hole physics \cite{Hawking:1982dh}. On the other hand, the SYK model is a one-dimensional, maximally chaotic quantum mechanical system describing the interactions of $N$ Majorana fermions, which can be solved exactly in the strong-coupling and Large $N$ limits.
 
Interestingly, the authors in \cite{Blommaert:2024ydx}-\cite{Blommaert:2024whf} generalized the two-dimensional gravity by introducing a periodic dilaton potential known as sine dilaton gravity (sDG), which acts as a ultraviolet (UV) completion of conventional JT gravity. The boundary dual of sDG is described by the double scaled SYK model (DSSYK) \cite{Lin:2022rbf}-\cite{Berkooz:2024lgq}. The DSSYK model features several intriguing properties, including the non-linear behavior of entropy, two distinct notions of temperature, and a bounded energy spectrum characterized by a periodic chord space Hamiltonian \cite{Lin:2022rbf}. On the other hand, the sDG model has proven useful in understanding two-dimensional quantum cosmological models \cite{Blommaert:2025avl}-\cite{Blommaert:2025rgw}. The sDG theory features an interpolating geometry capable of producing both de Sitter ($dS_2$) and Anti-de Sitter ($AdS_2$) space-time in different limits, depending on the choice of the principal branch \cite{Blommaert:2024ydx}. In other words, the DSSYK/sDG duality offers significant insights into gauge/gravity duality in lower dimensions and has consequently become a primary focus of recent literature.

It is worth noting that the authors in \cite{Bossi:2024ffa} explored one-loop corrections in the context of the DSSYK/sDG duality, while the authors in \cite{Aguilar-Gutierrez:2025pqp} explicitly computed the Krylov complexity in DSSYK and its bulk dual in sDG. Additionally, the authors in \cite{Mahapatra:2025fpx} computed the holographic central charges for sDG ($c_{\text{sDG}}$) and its low-energy limit, JT gravity ($c_{\text{JT}}$), finding that $c_{\text{sDG}} > c_{\text{JT}}$. This indicates that the sDG model can be interpreted as the UV completion of JT gravity. Furthermore, the authors in \cite{Mahapatra:2026aor} constructed a holographic $c$-function for the RG flow from sDG to JT gravity. Recently, irrelevant deformations in the context of DSSYK and their implications for sDG theory have been explored in \cite{Aguilar-Gutierrez:2026ogo}. It was demonstrated in \cite{Blommaert:2024ydx} that sDG can be rewritten as two copies of Liouville CFT at the classical level. In the present work, we discuss the effects due to marginal deformation into the sine dilaton gravity, following the approach of \cite{Giribet:2024men} and investigate its implications in the thermodynamics of marginally deformed 2D black holes.

Our analysis reveals that the deformation modifies key thermodynamic observables, such as the Hawking temperature, the Wald entropy, and the free energy of the black holes. Remarkably, these modifications preserve the standard form of the first law of black hole thermodynamics. A crucial physical implication of the marginal deformation is that it introduces a minimum temperature threshold ($T_0$). Below this critical value ($T < T_0$), the system enters an unphysical phase characterized by a fractional density of states and a negative Wald entropy. Above the critical value ($T > T_0$), the entropy becomes positive and increases with temperature, while the free energy decreases with a rise in temperature. Furthermore, we derive the corresponding Casimir function for deformed sDG (dsDG) which play key role in establishing the first law of black hole mechanics. Finally, we explore the infrared (IR) limit of deformed dsDG to obtain deformed JT (dJT) gravity, where we analyze its thermal properties and density of states. 

The rest of the paper is organized as follows. 
\begin{itemize}
             \item In Section 2, we discuss marginal deformations in the Liouville theory framework along the lines of \cite{Giribet:2024men}. We then translate the deformations into the sDG framework, thus setting up the background of our analysis.
             \item In Section 3, we compute the black hole solutions of the deformed sDG theory and comment on its conformal structure. 
             \item In Section 4, we study the thermodynamic properties of the deformed sDG black holes, investigating their phase stability and the density of states. 
             \item In Section 5, we compute the Komar integral and the associated Casimir function of the dsDG and verify the first law of black hole mechanics.
             \item In Section 6, we study the infrared (IR) limit of deformed sDG, which yields deformed JT (dJT) gravity. We compute the black hole solutions of dJT and examine their thermodynamic behavior.
             
\item Finally, in Section 7, we summarize our main conclusions and outline some interesting directions for future work.
         \end{itemize}

\section{Deformed sine dilaton gravity and Liouville CFT}
Marginal deformations in a Liouville conformal field theory (CFT) provide an interesting extension of the theory, as they preserve conformal invariance even at the level of conformal perturbation theory \cite{Giribet:2024men}. In other words, their study captures how the Liouville framework accommodates non-trivial modifications while retaining its underlying conformal structure. Furthermore, these deformations provide a controlled starting point for investigating how the interaction changes quantum observables, particularly the sphere partition function \cite{Giribet:2024men}.

Interestingly, sine dilaton gravity (sDG) \cite{Blommaert:2024ydx}, \cite{Mahapatra:2026aor} can also be recast into the Liouville theory framework. Therefore, it is naturally interesting to study the implications of these deformations on gravitational observables in the context of sDG and their interpretation in the corresponding double-scaled SYK (DSSYK) boundary theory. In this section, we establish the theoretical framework underlying our analysis by discussing the marginal deformations into the sDG framework \cite{Giribet:2024men}.

 The sDG is a theory of gravity in two space-time dimensions characterized by a sinusoidal potential \cite{Blommaert:2024ydx}. It serves as the UV completion of Jackiw-Teitelboim (JT) gravity, which features a linear potential \cite{Blommaert:2024ydx}-\cite{Blommaert:2024whf}. The complete action for the sDG theory\footnote{We include an overall minus sign in the action (\ref{eq:sDG_action}), defining the Euclidean partition function as $\mathcal{Z} = e^{-S_{\text{sDG}}}$. This convention aligns with \cite{Blommaert:2024ydx} to ensure path integral convergence and consistency with standard statistical mechanics.\label{fre0}} is given by \cite{Blommaert:2024ydx}
\begin{align}
S_{\rm sDG} = -\frac{1}{4|\log q|} \int d^2x \sqrt{g} \left[ \Phi R + 2\sin\Phi \right] + S_{\rm GHY} + S_{\rm ct},
\label{eq:sDG_action}
\end{align}
where $R$ represents the two-dimensional Ricci scalar, $\Phi$ denotes the dilaton scalar field, and the overall coupling $|\log q|$ is proportional to the inverse 2D gravitational constant. 

Furthermore, $S_{\rm GHY}$ corresponds to the Gibbons-Hawking-York boundary term [16], and $S_{\rm ct}$ represents the appropriate boundary counter-term, which are given below 
\begin{align}
S_{\rm GHY} = -\frac{1}{2|\log q|} \int_{\partial M} dt \sqrt{h} \, \Phi K, \quad
S_{\rm ct} = \frac{i}{2|\log q|} \int_{\partial M} dt \sqrt{h} \, e^{-i\Phi/2},
\label{eq:boundary_terms}
\end{align}
where $h$ denotes the determinant of the induced metric on the boundary $\partial M$, and $K$ represents the trace of the extrinsic curvature.




 Interestingly, the authors in \cite{Blommaert:2024ydx}, \cite{Mahapatra:2026aor}, show that in the conformal gauge (i.e., $g_{ab} = e^{2A}\hat{g}_{ab}$, $\hat{g}_{ab}$ being the flat metric), the sDG theory (\ref{eq:boundary_terms}) can be recast as a sum of two copies of the Liouville CFT
\begin{align}
    S_{\rm sDG} = \int d^2x \left( \frac{1}{4\pi} \left((\partial Z_1)^2 + 4\pi\mu_1 e^{b_1Z_1}\right) + \frac{1}{4\pi} \left((\partial Z_2)^2 + 4\pi\mu_2 e^{b_2Z_2}\right) \right), \label{lact1}
\end{align}
where $Z_1$ and $Z_2$ represent the redefined fields\footnote{Here, we use the following field redefinitions: $Z_1 = 2A + i\Phi$, and $Z_2 = -2iA-\Phi$. \label{fre1}}. Here, $b_1$ and $b_2$ are the corresponding Liouville parameters, and $\mu_1$, $\mu_2$ denote the cosmological constants.

In the literature, this theory (\ref{lact1}) is referred to as the Complex Liouville String (CLS) theory \cite{Giribet:2024men}, characterized by a sum of two Liouville CFTs with $b_2 = ib_1$ and complex conjugate central charges. The authors in \cite{Giribet:2024men} studied the CLS and its marginal deformation in detail within the Liouville theory framework. In particular, they introduced a marginal deformation of the following form
\begin{align}
\delta S = \lambda \int d^2x \sqrt{\hat{g}} \, O_a, \quad O_a = e^{2a_1 Z_1 + 2a_2 Z_2}
\label{operator}
\end{align}
where $O_a$ represents the deformation operator and $\lambda$ is the corresponding deformation parameter. Here, $a_1$ and $a_2$ are complex parameters constrained by the marginality condition \cite{Giribet:2024men}
\begin{align}
a_1(Q_1 - a_1) + a_2(Q_2 - a_2) = 1,
\label{marginality}
\end{align}
with $Q_i$ are the background charges, defined as $Q_i = b_i + b_i^{-1}$ for $i = 1, 2$. It can be shown that the marginality
condition (\ref{marginality}) ensures that the operator (\ref{operator}) possesses the precise scaling dimension required to be marginal\footnote{In a two-dimensional CFT, a scalar operator is marginal when its conformal weights satisfy $h = \bar{h} = 1$, such that its scaling dimension $\Delta = h + \bar{h} = 2$ precisely compensates the scaling weight of the integration measure $d^2x$.}. They proposed this deformation to study the Virasoro Minimal String \cite{Collier:2023cyw} and the Complex Liouville String in a unified framework, which has interesting connections with matrix models, de-Sitter gravity, and the DSSYK model \cite{Verlinde:2024zrh}-\cite{Narovlansky:2023lfz}.

In the following, we introduce a marginal deformation along the lines of \cite{Giribet:2024men} into the sDG model of 2D gravity (\ref{eq:sDG_action}), which serves as the foundation of our analysis. To begin with, we rescale the fields as
\begin{align}
    \Phi \rightarrow 2\pi b_1^2 \Phi\hspace{1mm},\hspace{2mm}
    Z_1 \rightarrow 2b_1Z_1\hspace{1mm},\hspace{2mm}
    Z_2 \rightarrow -2ib_2Z_2.\label{rfv2}
\end{align}
Using these rescalings (\ref{rfv2}), we can express $Z_1$ and $Z_2$ in terms of $A$ and $\Phi$ as follows (see footnote 2)
\begin{align}
    b_1 Z_1 = A + i\pi b_1^2\Phi\hspace{1mm},\hspace{2mm}
    b_2 Z_2 = A - i\pi b_1^2 \Phi.\label{expz2}
\end{align}

Next, we substitute the values of the background charges $Q_i$ in terms of $b_i$, along with the relation $b_2 = ib_1$, into condition (\ref{marginality}). This yields the following quadratic equation for $a_2$ in terms of $a_1$ and $b_1$
\begin{align}
    a_1
\left(
b_1 + \frac{1}{b_1} - a_1
\right)
+
a_2
\left[
i\left(b_1 - \frac{1}{b_1}\right) - a_2
\right]
=1.\label{quad_a2}
\end{align}

Solving the above equation (\ref{quad_a2}) for $a_2$, we obtain
\begin{align}
    a_2 = i(b_1 - a_1). \label{relation}
\end{align}
Thus, the marginality condition (\ref{marginality}) imposes the relation between $a_1$ and $a_2$ (\ref{relation}). Now substituting the expression for $Z_1$ and $Z_2$ from (\ref{expz2}) and the relation (\ref{relation}) into the operator (\ref{operator}), we find\footnote{In conformal gauge, the space-time metric can be expressed as $g_{ab} = e^{2A}\hat{g}_{ab}$, where $\hat{g}_{ab}$ is the flat space-time metric. In 2D theory, this yields $\sqrt{g} = e^{2A}\sqrt{\hat{g}}$.}
\begin{align}
\delta S = \lambda \int d^2x \sqrt{g} \, e^{2\pi i b_1(2a_1 - b_1)\Phi}.
\label{deformsDG}
\end{align}

In the subsequent analysis, we set the following condition on the Liouville parameters
\begin{align}
2\pi b_1(2a_1 - b_1) = 1.
\label{condition}
\end{align}
This constraint (\ref{condition}) further simplifies the deformation (\ref{deformsDG}) to
\begin{align}
\delta S = \lambda \int d^2x \sqrt{g} \, e^{i\Phi}.
\label{deformsDG2}
\end{align}

In the rest of the paper, we study this marginal deformation (\ref{deformsDG2}) in detail. Before proceeding further, it is worth noting that condition~(\ref{condition}) simplifies the deformation in such a way that it modifies the conformal factor while preserving the conformal $\text{AdS}_2$ structure (see Section~3.1 and \cite{Blommaert:2024ydx}). Moreover, this condition (\ref{condition}), along with the relation (\ref{relation}) and $b_2 =ib_1$, further restricts us to a special class of marginal deformations where the parameters obey the following conditions
\begin{align}
    a_1 = \frac{1}{2}\left(b_1 + \frac{1}{2\pi b_1}\right), \quad a_2 = \frac{i}{2}\left(b_1 - \frac{1}{2\pi b_1}\right). \label{eq:special_params}
\end{align}
These relations (\ref{eq:special_params}) completely fix the parameters of the deformation operator in terms of the underlying Liouville parameters, providing the stage for analyzing the physical implications in subsequent sections.


\section{Real deformed sDG}\label{Section real}
In this section, we study real marginally deformed sine dilaton gravity (rsDG) \cite{Giribet:2024men}. This deformation can be interpreted as the sum of a marginal deformation ($e^{i\Phi}$) and its complex conjugate ($e^{-i\Phi}$), yielding a real deformed sDG. For the single marginal deformation case, see Appendix A. Deforming the sDG leads to a modified space-time structure, which also alters the thermal properties of black holes. To be specific, we compute the black hole solutions for rsDG black holes in this section.

The action for rsDG \cite{Giribet:2024men} is given by
\begin{equation}
S=-\frac{1}{4|\log q|}\int d^2x \sqrt{g}\left[\Phi R+ 2\sin\Phi+2\lambda \cos\Phi\right]+S_{\rm GHY}+S_{\rm ct},\label{action}
\end{equation}
where $\lambda$ is a real deformation parameter. The associated boundary terms are given below
\begin{equation}
S_{\rm GHY}
=
-\frac{1}{2|\log q|}
\int_{\partial\mathcal M}
dt \sqrt{h}\,\Phi K\hspace{1mm},\hspace{2mm}S_{ct}=\frac{\alpha}{2|\log q|}
\int_{\partial M}
d\tau\,
\sqrt{h}\, \label{GHY_ct}
e^{-i\Phi/2},
\end{equation}
where $\alpha = e^{i\pi/4}\sqrt{i + \lambda}$ is a constant chosen to cure all boundary divergences. 

\subsection{Black hole solutions}
In this subsection, we examine the dynamical equations of motion associated with rsDG \eqref{action}. 
In particular, we obtain the black hole solutions in the static gauge \cite{schrodinger1950}. On varying the action (\ref{action}) with respect to 
the space-time metric ($g_{\mu\nu}$) and the dilaton ($\Phi$), we obtain the following equations of motion:
\begin{align}
\nabla_\mu\nabla_\nu\Phi-g_{\mu\nu}\Box\Phi-\frac12 g_{\mu\nu}\mathcal{V}(\Phi)=0,\label{eq1}\\
R - \mathcal{V}'(\Phi)=0,\label{eq2}
\end{align}
where the potential is defined as $\mathcal{V}(\Phi) = -2 \sin \Phi - 2 \lambda \cos \Phi$. Here, $'$ denotes the derivative with respect to the dilaton ($\Phi$).

Now, one can easily solve the field equations \eqref{eq1} and \eqref{eq2} in the Euclidean static gauge \cite{schrodinger1950}
\begin{align}
ds^2 = F(r) \, dt^2 + \frac{dr^2}{F(r)}, \label{gauge}
\end{align}
where $r$ is the radial coordinate. Solving equations \eqref{eq1} and \eqref{eq2} in the gauge \eqref{gauge} yields the following deformed black hole solutions\footnote{Notice that although the calculations are performed in the Euclidean signature, the corresponding Lorentzian signature metric, $ds^2 = -F(r)dt^2 + \frac{dr^2}{F(r)}$, yields identical black hole solutions as in (\ref{bhsol1})-(\ref{bhsol2}).}
\begin{align}
F_{\text{bh}}(r) &= 2 \cos \theta - 2 \cos r + 2 \lambda \sin r, \label{bhsol1} \\
\Phi_{\text{bh}}(r) &= r, \label{bhsol2}
\end{align}
where the subscript ``$\text{bh}$'' denotes the black hole solution, and $\theta$ is an integration constant. 

The black hole horizon $r_h$ can be obtained by solving $F_{\text{bh}}(r_h) = 0$, which yields
\begin{align}
  r_h = \cos^{-1}\left(\frac{\cos \theta}{\sqrt{1+\lambda^2}}\right) - \tan^{-1}\lambda.\label{horizon}
\end{align}
Similarly, the boundary of spacetime is located at $r_{\text{bdy}} = \frac{\pi}{2} - \tan^{-1}\lambda + i\infty$, where $F(r)$ diverges. Taking the limit $\lambda \to 0$, we simultaneously recover the undeformed horizon radius $r_h = \theta$ and the undeformed boundary $r_{\text{bdy}} = \frac{\pi}{2} + i\infty$ \cite{Blommaert:2024ydx}. In other words, the deformation parameter $\lambda$ modifies both the black hole horizon and the boundary structure.

Remarkably, under the following coordinate transformation
\begin{align}
  r = \frac{\pi}{2} - \tan^{-1}\lambda + i \log \left( \frac{\rho + i \cos\theta}{\sqrt{1+\lambda^2}} \right), \label{coord_trans}
\end{align}
the static line element \eqref{gauge} can be re-expressed in the form of a conformally $\text{AdS}_2$ metric,
\begin{align}
  ds^2 = \frac{e^{i \tilde{r}}}{\sqrt{1+\lambda^2}} \left[ (\rho^2 - \rho_h^2) d\tau^2 + \frac{d\rho^2}{\rho^2 - \rho_h^2} \right]=\frac{e^{i \Phi(\tilde{r})}}{\sqrt{1 + \lambda^2}} \, ds^2_{\text{eff}}, \label{conf_ads2}
\end{align}
where $\tilde{r} = r + \tan^{-1}\lambda$ and $\rho_h = \sqrt{\sin^2\theta + \lambda^2}$. Here, $ds^2_{\text{eff}}$ stands for the effective $\text{AdS}_2$ metric, given below
\begin{align}
    ds^2_{\text{eff}}=(\rho^2 - \rho_h^2) d\tau^2 + \frac{d\rho^2}{\rho^2 - \rho_h^2}.
\end{align}

Notice that our choice of the radial contour~\eqref{coord_trans} is similar to that of the undeformed theory \cite{Blommaert:2024ydx}, with the addition of the deformation parameter $\lambda$. For the undeformed theory, this choice demonstrated that a physical bulk probe experiences an effective $\text{AdS}_2$ geometry, which was crucial for matching correlation functions and establishing the sDG/DSSYK duality~\cite{Blommaert:2024ydx}-\cite{Blommaert:2024whf}. In a similar spirit, we argue that for the deformed theory, a bulk probe consisting of a massive scalar field non-minimally coupled to the dilaton,
\begin{equation}
\int d^2x \sqrt{g} \left( g^{\mu \nu} \partial_\mu \phi \partial_\nu \phi + m^2 \sqrt{1 + \lambda^2} \, e^{-i \Phi(\tilde{r})} \phi^2 \right) = \int d^2x \sqrt{g_{\text{eff}}} \left( g^{\mu \nu}_{\text{eff}} \, \partial_\mu \phi \partial_\nu \phi + m^2 \phi^2 \right), \label{scalar_action}
\end{equation}
will likewise experience an effective $\text{AdS}_2$ geometry, thereby preserving the structure of the holographic dictionary.

\section{Thermodynamics and phase stability}\label{sec_thermo_rsDG}
In this section, we investigate the thermodynamic properties of the real deformed sine dilaton gravity (rsDG) black holes (\ref{bhsol1})-(\ref{bhsol2}). In particular, we obtain their Hawking temperature \cite{Hawking:1975vcx} and examine thermal observables such as Wald entropy \cite{Wald:1993nt}-\cite{Lala:2020lge} and free energy \cite{Gibbons:1976ue}-\cite{Rathi:2021mla}. Finally, we analyze the phase stability of these black holes.

In literature, the Hawking temperature ($T_H$) \cite{Hawking:1975vcx} is defined as
\begin{align}
T_H=\frac{F'_{\text{bh}}(r_h)}{{4\pi}}.\label{Hawking_temp}
\end{align}

For rsDG balck holes, the Hawking temperature can be obtained by differentiating equation \eqref{bhsol1} with respect to radial coordinate $r$ at the horizon $r = r_h$, which yields the following
\begin{align}
    T_H = \frac{\sin r_h + \lambda\cos r_h}{2\pi}.\label{hwtemp1}
\end{align}
On the other hand, evaluating \eqref{bhsol1} directly at $r = r_h$ gives the horizon condition
\begin{align}
    \cos r_h - \lambda \sin r_h = \cos \theta. \label{horizoncdn}
\end{align}

Notice that one can express the Hawking temperature (\ref{hwtemp1}) in a more compact form. To this end, squaring and adding equations (\ref{hwtemp1}) and (\ref{horizoncdn}), followed by a straightforward rearrangement of terms, leads to
\begin{align}
    T_H = \frac{1}{2\pi}\sqrt{\lambda^2 + \sin^2 \theta}. \label{hwtemp2}
\end{align}
As expected, in the undeformed limit ($\lambda \rightarrow 0$), the equation \eqref{hwtemp2} smoothly reduces to the standard Hawking temperature of the undeformed theory \cite{Blommaert:2024ydx}.

Now, we compute the Wald entropy ($S_W$) \cite{Wald:1993nt} associated with the rsDG black holes. We obtain $S_W$ by varying the action \eqref{action} with respect to the Riemann curvature tensor ($R_{\mu\nu\alpha\beta}$) as follows
\begin{align}
  S_W = -2\pi \frac{\partial L}{\partial R_{\mu\nu\alpha\beta}}\Bigg|_{r=r_h} \epsilon_{\mu\nu} \epsilon_{\alpha\beta},\label{wald}
\end{align}
where $L$ is the Lagrangian and $\epsilon_{\mu\nu}$ is the anti-symmetric tensor, with a normalization condition given as $\epsilon^{\mu\nu}\epsilon_{\mu\nu}=-2$ \cite{Wald:1993nt}.

Using equations \eqref{action} and \eqref{wald}, we obtain
\begin{align}
S_{\text{W}} = \frac{\pi \Phi(r_h)}{|\log q|}=\frac{\pi r_h}{|\log q|}=\frac{\pi}{|\log q|} \left[ \cos^{-1}\left( \frac{\cos \theta}{\sqrt{1+\lambda^2}} \right) - \tan^{-1}\lambda \right], \label{wald_result}
\end{align}
where we have used the dilaton profile \eqref{bhsol2} and the horizon radius $r_h$ given in \eqref{horizon}. This demonstrates that the Wald entropy of the black holes explicitly depends on the deformation parameter $(\lambda)$.

Next, we examine the free energy ($\mathcal{F}_{\text{bh}}$) associated with the deformed black hole solution \eqref{bhsol1}. Within the Euclidean path integral formulation, the free energy \cite{Gibbons:1976ue} is defined as
\begin{align}
\mathcal{F}_{\text{bh}} = -\beta^{-1} \log Z, \label{free_energy_def}
\end{align}
where $Z = e^{-S_{\text{on-shell}}}$ denotes the thermodynamic partition function, $S_{\text{on-shell}}$ is the corresponding Euclidean on-shell action, and $\beta = T_H^{-1}$ is the inverse Hawking temperature.

Notice that the free energy receives contributions from both bulk and boundary terms. To determine the bulk contribution, we first compute the Ricci scalar, $R = -2 \cos r + 2 \lambda \sin r$. Substituting $R$ along with the solutions \eqref{bhsol1}-\eqref{bhsol2} into equation \eqref{action}, and integrating over the radial and time coordinates, we obtain
\begin{align}
    S_{\text{bulk}} = -\frac{\beta}{4|\log q|}\left[\mathcal{B}(r) \right]^{r_c}_{r_h},\label{bulk_cntr}
\end{align}
where $\mathcal{B}(r) = -2r\sin r - 4 \cos r -2\lambda r \cos r + 4 \lambda \sin r$ and $r_c$ is the boundary cut-off. In the asymptotic limit, $r_c\rightarrow\pi/2 + i y_c$ and $y_c\rightarrow \infty$.

Similarly, the contributions from the Gibbons-Hawking-York (GHY) boundary term and the counter-terms are given by
\begin{align}
    S_{\text{GHY}}=&\hspace{1mm} -\frac{\beta}{2|\log q|}\left[ r_c\sin r_c + \lambda r_c \cos r_c\right].\label{GHY_cntr}\\
    S_{\text{ct}}=&\hspace{1mm}\frac{\beta}{2|\log q|} (i+\lambda)e^{y_c}
+\frac{\beta}{2|\log q|}\cos\theta.\label{action_ct}
\end{align}



    Adding the bulk contribution (\ref{bulk_cntr}) and the boundary contributions (\ref{GHY_cntr})-(\ref{action_ct}), we obtain the renormalized on-shell action ($S_{\text{on-shell}}$) as follows
\begin{align}
    S_{\text{on-shell}}= -\frac{\beta}{2|\log q|} \left[ r_h \sin r_h + \cos r_h + \lambda r_h \cos r_h - \lambda \sin r_h \right].\label{v2onshellaction}
\end{align}

Using the above expression (\ref{v2onshellaction}) in the definition (\ref{free_energy_def}), we obtain the free energy as follows
\begin{align}
\mathcal{F}_{\text{bh}} = -\frac{1}{2|\log q|} \left[ r_h \sin r_h + \cos r_h + \lambda r_h \cos r_h - \lambda \sin r_h \right]. \label{free_energy_res}
\end{align}

Notice that in the limit $\lambda \to 0$, we smoothly recover the black hole free energy corresponding to pure sine dilaton gravity \cite{Blommaert:2024ydx}. In other words, the deformation parameter $\lambda$ directly shifts both the thermodynamic entropy and the total free energy of the black hole, demonstrating how the marginal deformations alter the thermal stability.

In the following, we express the Wald entropy \eqref{wald_result} and free energy \eqref{free_energy_res} of the deformed black holes explicitly in terms of the Hawking temperature \eqref{Hawking_temp}. This formulation enables us to analyze their temperature dependence for various values of the deformation parameter $(\lambda)$. To begin with, we express $r_h$ in terms of the Hawking temperature. Toward this end, we solve equations (\ref{hwtemp1}) and (\ref{horizoncdn}) for $\sin r_h$ and $\cos r_h$, which yields
\begin{align}
   \sin r_h = \frac{2\pi T - \lambda \cos \theta}{1 + \lambda^2}\hspace{1mm},\hspace{2mm}
    \cos r_h = \frac{2 \pi \lambda T + \cos \theta}{1 + \lambda^2}.\label{cosrh}
\end{align}
Here, we drop the subscript $H$ in temperature for brevity.

It is straightforward to determine $\cos\th$ from equation (\ref{cosrh}) using the trigonometric identity $\sin^2 r_h + \cos^2 r_h = 1$, which yields
\begin{align}
    \cos \theta = \sqrt{1+ \lambda^2 - 4 \pi^2T^2}. \label{theta}
\end{align}
Substituting equation (\ref{theta}) back into (\ref{cosrh}), we obtain

\begin{align}
    \sin r_h = \frac{2\pi T - \lambda\sqrt{1+\lambda^2 -4\pi^2 T^2}}{1+\lambda^2}\hspace{1mm}, \hspace{2mm}
    \cos r_h = \frac{2\pi\lambda T + \sqrt{1+\lambda^2-4\pi^2 T^2}}{1+\lambda^2} \label{cosrh2}
\end{align}

Now we define the quantities, $\alpha \equiv \sin^{-1} \left( \frac{2\pi T}{1+\lambda^2}\right)$ and $\delta \equiv\tan^{-1} \lambda$, such that we have
\begin{align}
\sin \alpha = \frac{2\pi T}{\sqrt{1+\lambda^2}}\hspace{1mm},\hspace{2mm}
\cos \alpha = \frac{\sqrt{1+\lambda^2-4\pi^2T^2}}{\sqrt{1+\lambda^2}}\hspace{1mm},\hspace{2mm} 
    \cos \delta = \frac{1}{\sqrt{1+\lambda^2}}\hspace{1mm}, \hspace{2mm} \sin \delta = \frac{\lambda}{\sqrt{1+\lambda^2}}\label{cosdelta}
\end{align}
Substituting the components of equation (\ref{cosdelta}) into equation (\ref{cosrh2}) for $\sin r_h$, and applying the trigonometric subtraction identity for sine, we obtain
\begin{align}
    \sin r_h = \sin \alpha \cos \delta - \cos \alpha \sin \delta = \sin (\alpha - \delta).
\end{align}

Selecting the appropriate physical branch for $r_h$, we arrive at 
\begin{align}
r_h(T) = \alpha - \delta = \sin^{-1}\left(\frac{2\pi T}{\sqrt{1+\lambda^2}}\right) - \tan^{-1}\lambda.\label{rht}
\end{align}

Using the above expression \eqref{rht}, one can express the Wald entropy \eqref{wald_result} in terms of the temperature as follows
\begin{align}
    S_{\rm W}(T)=\frac\pi{|\log  q|}\left[\sin^{-1}\!\left(\frac{2\pi T}{\sqrt{1+\lambda^2}}\right)-\tan^{-1}\lambda
\right].\label{ET}
\end{align}
In order to express the free energy (\ref{free_energy_res}) as a function of the temperature $T$, we first substitute equation (\ref{hwtemp1}) into the free energy to replace the combination $\sin r_h + \lambda \cos r_h$ with $2\pi T$ which yields
\begin{align}
    \mathcal{F}_{\text{bh}} =  -\frac{1}{2|\log q|}\left[ 2\pi T r_h(T) + \cos r_h - \lambda \sin r_h\right].\label{free_energy_intm}
\end{align}

Next, substituting the expressions from equations (\ref{horizoncdn}), (\ref{theta}), and (\ref{rht}) into equation (\ref{free_energy_intm}) yields
\begin{align}
   \mathcal{F}_{\text{bh}}(T) =-\frac{1}{2|\log q|}
\left[2\pi T\left(\sin^{-1}\left(\!\frac{2\pi T}{\sqrt{1+\lambda^2}}\right)-\tan^{-1}\lambda
\right)+\sqrt{1+\lambda^2-4\pi^2T^2}\right].\label{FT}
\end{align}

Before proceeding further, it is worth noting that the entropy \eqref{ET} and free energy \eqref{FT} satisfy the fundamental thermodynamic relation
\begin{align}
\frac{d \mathcal{F}_{\rm bh}(T)}{dT} = -S_{\rm W}(T).
\end{align}

\begin{figure}[htp]
    \centering
    \begin{subfigure}[b]{0.48\textwidth}
        \centering
        \includegraphics[width=\textwidth]{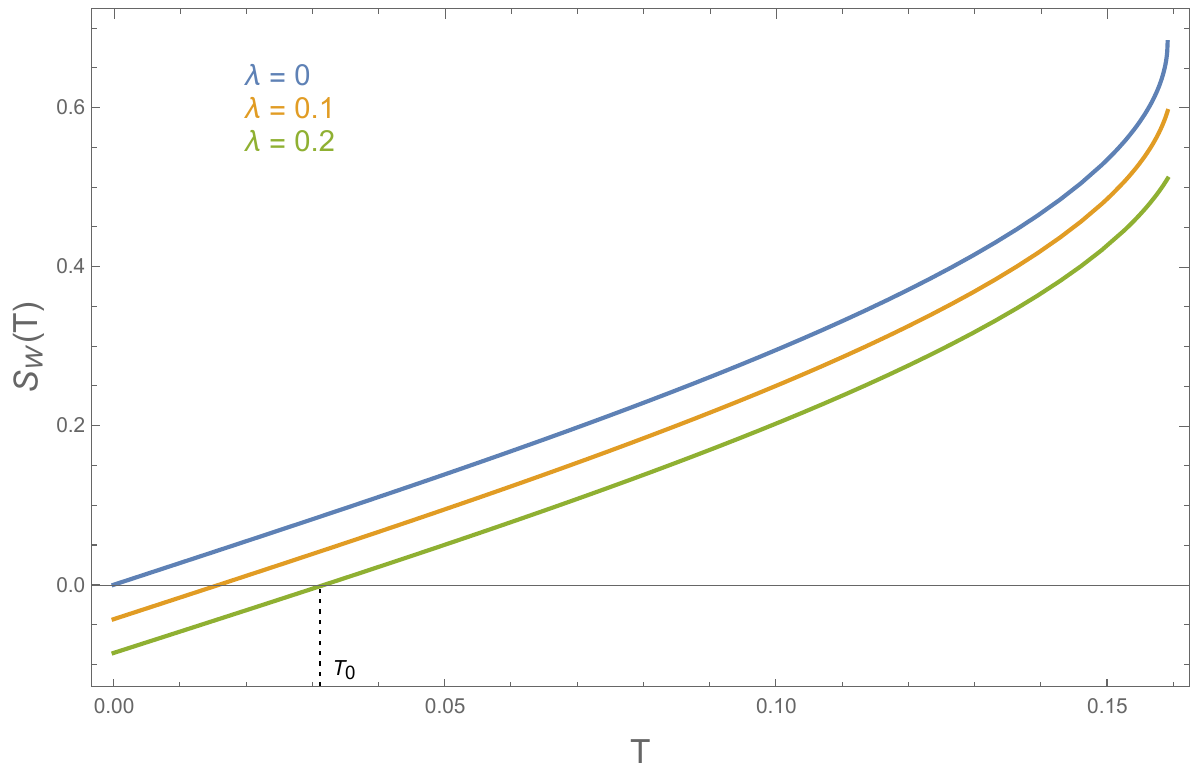}
        \caption{Entropy vs temperature}
        \label{Eimg}
    \end{subfigure}
    \hfill
    \begin{subfigure}[b]{0.48\textwidth}
        \centering
        \includegraphics[width=\textwidth]{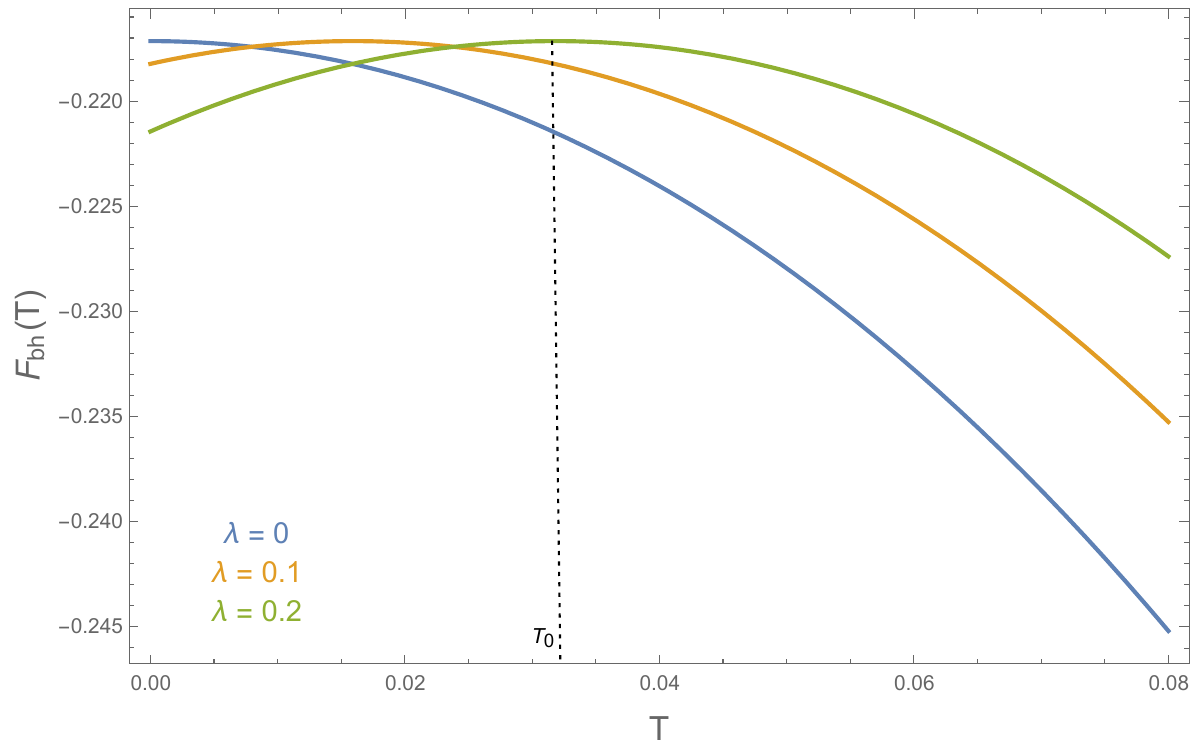}
        \caption{Free energy vs temperature}
        \label{Fimg}
    \end{subfigure}
    \caption{Phase stability of rsDG black holes for coupling parameter, $q=0.1$.}
    \label{fig:entropy_free_energy}
\end{figure}

Next, we plot these quantities \eqref{ET}-\eqref{FT} against the temperature for various values of $\lambda$. It is interesting to note that the deformation parameter ($\lambda$) significantly alters both the entropy and the free energy of the black holes. For instance, in the absence of deformation ($\lambda = 0$), the entropy starts at zero and grows continuously with temperature, as shown in Figure~\ref{Eimg}. However, in the presence of deformation, there exists a minimum temperature $T_0$ below which the entropy becomes negative and corresponds to unphysical (micro) states or degree of freedom in sine dilaton gravity.

\par The critical temperature $T_0$ entirely depends on the deformation parameter ($\lambda$) and can be obtained analytically via the vanishing condition of the Wald entropy \eqref{wald_result}, as shown below
\begin{align}
S_{\rm W}(T_0) = 0 \implies T_0 = \frac{\lambda}{2\pi}.\label{crict}
\end{align}

Physically, the thermodynamic entropy provides a measure of the density of states ($\Omega$) i.e., $S = \log \Omega$. In the temperature regime ($T < T_0$), one finds that
\begin{align}
    \sin^{-1} \left( \frac{2\pi T}{\sqrt{1+\lambda^2}} \right) <  \sin^{-1}\left(\frac{2\pi T_0}{\sqrt{1+\lambda^2}} \right)=\sin^{-1} \left( \frac{\lambda}{\sqrt{1+\lambda^2}} \right)=\tan^{-1}\lambda.\label{negrh}
\end{align}
This implies that $r_h(T)$ given by (\ref{rht}) is negative in the region $T<T_0$. As a result, the Wald entropy $(S_W)$ (\ref{wald_result}), which is proportional to the horizon value ($r_h$), turns negative. This formally leads to a density of states ($\Omega(T)$) strictly less than unity as shown below
\begin{align}
    \Omega(T) = \exp(S_{\rm W}) = \exp\left(\frac{\pi\Delta}{|\log q|} \right), \label{eq:density_states}
\end{align}
where $\Delta =  \sin^{-1}\left( \frac{2\pi T}{\sqrt{1+\lambda^2}} \right) - \tan^{-1}\lambda $.  

Now, we perform a perturbative expansion of $\Omega(T)$ in \eqref{eq:density_states} around the critical temperature $T_0$. Defining $T = T_0 - \epsilon$, where $0 < \epsilon \ll 1$ denotes a small deviation below the critical threshold, the density of states expands as
\begin{align}
\Omega(T_0 - \epsilon) \approx 1 - \frac{2\pi^2}{\vert \log q \vert} \, \epsilon + \mathcal{O}(\epsilon^2),\label{omegaexpan}
\end{align}
which is strictly less than unity for any non-vanishing deformation ($\epsilon > 0$).

The condition $\Omega(T) < 1$ in \eqref{omegaexpan} signals a fundamental breakdown of the standard semiclassical thermodynamic description, as a well defined microcanonical ensemble strictly requires $\Omega(T) > 1$ (and corresponding positive entropy $S_{\rm W} > 0$). This unphysical regime ($T < T_0$) originates directly from the real marginal deformation parameter $\lambda$. In two dimensional dilaton gravity, the dilaton field $(\Phi)$ determines the effective gravitational coupling via $G_{\rm eff} \sim \Phi^{-1}$ \cite{Almheiri:2014cka}. Below the critical temperature $(T<T_0)$, the deformation forces the horizon radius $r_h(T)$ (\ref{rht}) and consequently the dilaton at the horizon $\Phi(r_h)$ (\ref{bhsol2}) into negative values (\ref{negrh}). This renders the effective gravitational coupling negative, indicating an inaccessible region where the physical degrees of freedom are suppressed.

Notice that the exclusion of this low-temperature branch is a direct consequence of requiring a positive effective gravitational coupling. Although the density of states $\Omega(T)<1$ in this regime, it remains positive and non-vanishing, unlike the negative spectral densities encountered in certain semiclassical pure-gravity models \cite{Maloney:2007ud}-\cite{Benjamin:2019stq}. Consequently, this classical unphysical branch does not, by itself, imply that the marginal deformation renders the boundary theory non-unitary \cite{Cardy:1986ie}. 


Similarly, the plot of the free energy (Figure~\ref{Fimg}) versus temperature reveals a local maximum at the critical temperature $T_0$. For lower temperatures ($T < T_0$), $\mathcal{F}_{\rm bh}$ increases monotonically with temperature until it reaches a local peak at $T = T_0$. Beyond this critical threshold, $\mathcal{F}_{\rm bh}$ turns downward and decreases monotonically as the temperature continues to rise, leading to a physically stable black hole phase. Furthermore, larger deformations cause this local maximum in $\mathcal{F}_{\rm bh}$ to shift toward higher temperatures (\ref{crict}). In the undeformed limit ($\lambda \to 0$), this local maximum disappears entirely, leaving a free energy profile that decreases monotonically across all temperatures.

To summarize, Figure~\ref{fig:entropy_free_energy} demonstrates that the presence of deformation gives rise to an unphysical phase at temperatures below the critical value ($T < T_0$). In this regime, the entropy is negative and the free energy increases monotonically with rising temperature. Once the temperature exceeds the critical threshold ($T > T_0$), we obtain a physical black hole phase characterized by positive entropy and negative free energy.

\section{Komar integral and black hole thermodynamics}
In this section, we compute the Komar integral \cite{Komar:1958wp} and verify the first law of thermodynamics \cite{Bardeen:1973gs} for real marginally deformed black holes. It is worth noting that the standard construction of conserved charges using the Komar integral does not work for sine-dilaton gravity. Therefore, we employ the Grumiller--Kummer--Vassilevich (GKV) formalism \cite{Grumiller:2021cwg}-\cite{Grumiller:2002nm} to obtain the Casimir function for the real deformed sine dilaton gravity (rsDG) black holes. One of the advantages of this formalism is that the Casimir function is a
coordinate independent conserved quantity which, for black-hole solutions,
is naturally identified with the conserved mass.

In order to compute the Casimir function, we begin with the action for canonical two-dimensional dilaton gravity containing at most second-order derivatives, given by \cite{Grumiller:2002nm}
\begin{equation}
I
=
-\frac{\kappa}{4\pi}
\int d^2x\,
\sqrt{-g}
\left[
XR
-
U(X)(\nabla X)^2
-
2V(X)
\right],
\label{canonical_action}
\end{equation}
where $R$ is the two-dimensional Ricci scalar, $X$ denotes the dilaton, $U(X)$ represents the kinetic coupling, $V(X)$ is the dilaton potential, and $\kappa$ is an overall coupling constant.

Notice that our real marginally deformed sine-dilaton gravity theory \eqref{action} exhibits a similar structure. By comparing the canonical action \eqref{canonical_action} with our rsDG theory \eqref{action}, we establish the following mappings\footnote{Here, the overall coupling constant is identified as $\kappa = \frac{\pi}{|\log q|}$.}
\begin{align}
    X = \Phi, \quad U(X) = 0, \quad V(X) = -\sin X - \lambda \cos X. \label{mapping}
\end{align}
Consequently, the rsDG belongs to the class of dilaton gravity models characterized by the absence of kinetic dilaton coupling.

Following the Grumiller--Kummer--Vassilevich (GKV) formalism, the conserved Casimir function ($C$) \cite{Grumiller:2021cwg} is defined as
\begin{equation}
    C = e^{Q(X)} X^+ X^- + w(X), \label{general_casimir}
\end{equation}
where $X^+$ and $X^-$ denote the light-cone coordinates \cite{Grumiller:2002nm}. Here, the auxiliary functions $Q(X)$ and $w(X)$ are given by
\begin{align}
    Q(X) = \int^{X} U(y) \, dy, \quad w(X) = -\int^{X} e^{Q(y)} V(y) \, dy. \label{auxdef}
\end{align}

Substituting the mapping \eqref{mapping} into the definitions \eqref{auxdef} and performing the explicit integrations, we find that the auxiliary functions simplify to
\begin{align}
    Q(X) = 0, \qquad w(X) = -\cos X + \lambda \sin X. \label{auxfunc}
\end{align}
Again, substituting these auxiliary functions into \eqref{general_casimir} yields the Casimir function for real deformed sine-dilaton gravity as follows
\begin{equation}
    C = -\frac{1}{2} (\nabla X)^2 - \cos X + \lambda \sin X, \label{casimir_rsDG}
\end{equation}
where we have utilized the geometric relation\footnote{For detailed calculation, see appendix B.} $X^+ X^- = -\frac{1}{2} (\nabla X)^2$ \cite{Grumiller:2002nm}. 

Notice that the addition of the marginal deformation parameter $(\lambda)$ directly modifies the scalar potential $V(X)$, while leaving the kinetic coupling vanishing ($U(X) = 0$). As a result, the exterior derivative of the Casimir function continues to vanish on-shell\footnote{For detailed calculation, see appendix C.}, i.e., $dC = 0$ \cite{Grumiller:2002nm}. Furthermore, \(\lambda\) acts solely as a target-space coupling and introduces no explicit coordinate dependence. Nevertheless, it smoothly deforms the background geometry and the thermodynamic properties of black holes while preserving the exact integrability of the rsDG field equations and the underlying Poisson-algebraic conservation structure of the theory \cite{Grumiller:2002nm}.

In order to simplify the Casimir function (\ref{casimir_rsDG}), we first expand the kinetic term $(\nabla X)^2$ as follows
\begin{align}
    (\nabla X)^2 = g^{rr} \partial_rX\partial_rX.\label{gradx0}
\end{align}
Now using the mapping (\ref{mapping}) and black hole solutions (\ref{bhsol1})-(\ref{bhsol2}) into  (\ref{gradx0}), we obtain
\begin{align}
    (\nabla X)^2= 2 \cos \theta - 2\cos r + 2\lambda\sin r. \label{gradx}
\end{align}
Finally substituting (\ref{gradx}) back into the Casimir function (\ref{casimir_rsDG}), we obtain
\begin{align}
    C = -\frac{1}{2}(2\cos\theta - 2\cos r +2 \lambda \sin r) - \cos r+\lambda \sin r = -\cos\theta.\label{ci}
\end{align}
Notably, the deformation parameter $\lambda$ does not explicitly appear in the expression for the Casimir invariant, as expected.

In literature \cite{Grumiller:2021cwg}-\cite{Grumiller:2002nm}, the mass of black hole is defined to be the Casimir invariant up to appropriate normalization. Along the same line, we define the mass of the rsDG black hole as follows
\begin{equation}
    M =\frac{C}{2|\log q|}= -\frac{\cos\theta}{2|\log q|}\implies dM = \frac{\sin \theta}{2 |\log q|}d\theta, \label{mass_rsDG}
\end{equation}
where the overall normalization constant is fixed by requiring that the first law of black-hole mechanics is satisfied identically. This result is consistent with the Arnowitt--Deser--Misner (ADM) energy (or conserved mass measured at spatial infinity) previously obtained for standard sDG as well as the DSSYK result \cite{Blommaert:2024ydx}. Although the mass $M$ is negative, it remains bounded from below, suggesting that a physical (positive) mass can be obtained via an appropriate shift in the energy spectrum, as discussed below.

The negative sign of the ADM energy ($E(\theta)$) (\ref{mass_rsDG}) in sine dilaton gravity,
\begin{align}
E(\theta)
=
-\frac{\cos\theta}{2|\log q|},\label{edam1}
\end{align}
can be naturally understood from the bounded DSSYK spectrum. The
spectral parameter $\theta$ ranges within $0\leq\theta\leq\pi,$ bounding the energy as \cite{Okuyama:2025hsd}
\begin{align}
-\frac{1}{2|\log q|}
\leq
E(\theta)
\leq
\frac{1}{2|\log q|}.
\end{align}

Notice that the spectrum is symmetric around $E=0$ (occurring at $\theta = \pi/2$), while $\theta=0$ and $\theta=\pi$ correspond to the
lower and upper spectral edges \cite{Blommaert:2024ydx}, respectively. The entropy is maximum at $\theta=0$ and degenerate for all other values of $\theta$ \cite{Blommaert:2024ydx}. For every value of the entropy, there exists two corresponding values of $\theta$, one in the range $0\leq\theta<\pi/2$ and the other in the range $\pi/2 <\theta \leq \pi$. Therefore, to avoid over counting of states, it is important to consider only one of the ranges which is a counterpart of the other. In other words, the opposite sign of the cosine parametrization can be obtained by
the reparametrization $\theta\rightarrow\pi-\theta,$ under which $\cos(\pi-\theta)=-\cos\theta$ \cite{Okuyama:2025hsd}. This transformation exchanges the two endpoints of the bounded
spectral interval and reverses the sign of the energy parametrization.

Following the convention of the authors in \cite{Blommaert:2024ydx}, the low-energy
sector corresponds to the lower edge. Expanding the energy around this endpoint ($\theta=0$) yields
\begin{align}
E(\theta)
=
-\frac{1}{2|\log q|}
+
\frac{\theta^2}{4|\log q|}
+
\mathcal{O}(\theta^4),
\end{align}
implying a minimum energy of the spectrum given by
\begin{align}
E_{\min}
=
-\frac{1}{2|\log q|}.
\end{align}

Next, we define the excitation energy ($E_{\rm exc}$) of the black holes relative to the lower spectral edge, as $E_{\rm exc}=
E-E_{\min},$ we find
\begin{align}
E_{\rm exc}
=
\frac{1-\cos\theta}{2|\log q|}
\simeq
\frac{\theta^2}{4|\log q|}>0. \label{peadm}
\end{align}
where we have used equation (\ref{edam1}). This positive excitation energy (\ref{peadm}) exhibits the quadratic dependence
characteristic of the Schwarzian regime. Thus, the negative ADM energy
of sDG is naturally interpreted as the absolute energy of a state
within the bounded DSSYK spectrum, whereas the positive Schwarzian
energy is obtained by measuring the energy relative to the lower
spectral edge. Consequently, the sign of the ADM energy is associated with the parametrization of the black hole horizon, chosen such that the lower edge of the spectrum yields the correct Schwarzian energy \cite{Blommaert:2024ydx}.

 Since the deformation parameter does not affect the Casimir invariant or the conserved mass, it leaves the first law of thermodynamics unchanged. As discussed previously, the first law of black hole thermodynamics relates the variation of the conserved mass ($M$) to the Hawking temperature ($T_H$) and the Wald entropy ($S_{\text{W}}$) via \cite{Bardeen:1973gs}
\begin{align}
dM = T_H \, dS_{\text{W}}.
\label{first_law}
\end{align}
For real deformed sine dilaton gravity, the Hawking temperature \eqref{Hawking_temp} and the variation of the Wald entropy \eqref{wald_result} are given by
\begin{align}
T_H = \frac{1}{2\pi} \sqrt{\lambda^2 + \sin^2 \theta}, \qquad 
dS_{\text{W}} = \frac{\pi}{|\log q|} \frac{\sin \theta}{\sqrt{\lambda^2 + \sin^2 \theta}} \, d\theta.
\end{align}

Taking the product $T_H \, dS_{\text{W}}$ directly yields
\begin{align}
T_H \, dS_{\text{W}} = \frac{\sin \theta}{2|\log q|} \, d\theta=dM,
\end{align}
which precisely reproduces the first law of black hole mechanics. It is worth noting that, although the marginal deformation $\lambda$ modifies both the temperature and the entropy differential individually, their non-trivial $\lambda$ dependence cancels out exactly in their product.

\section{Deformed JT gravity}\label{sectiondjt}
In this section, we examine the infrared (IR) or low energy limit of real marginally deformed sine dilaton gravity (rsDG), defined by the action \eqref{action}. This IR regime is realized by appropriately rescaling both the dilaton field $\Phi$ and the marginal deformation parameter $\lambda$, followed by taking the limit $|\log q| \to 0$. Within this limiting regime, we construct the corresponding black hole solutions and analyze their thermodynamic properties.

To formulate the low energy limit, we begin by redefining the dilaton field $\Phi$ and the deformation parameter $\lambda$ as follows
\begin{align}
    \Phi \rightarrow 2|\log q| \overline{\Phi}, \qquad \lambda \rightarrow 2|\log q| \overline{\lambda}, \label{fieldredef}
\end{align}
where the overline represents the redefine quantities.

Substituting these (\ref{fieldredef}) redefinitions into the action \eqref{action} and taking the limit $|\log q| \to 0$, we arrive at\footnote{One can also obtain the low energy limit of the action \eqref{action} by expanding around $\Phi = r = 0$ and keeping only the leading term in $\Phi$ \cite{Mahapatra:2026aor}. Additionally, one should keep the constant $\theta$ sufficiently small in the black hole solutions \eqref{bhsol1}-\eqref{bhsol2}.}
\begin{align}
    S_{\text{dJT}} = -\frac{1}{2} \int d^2x \sqrt{g} \left[ \overline{\Phi}(R+2) + 2\overline{\lambda} \right] + S_{\text{GHY}} + S_{\text{ct}}.\label{jtact}
\end{align}
Notice that this action (\ref{jtact}) corresponds to the standard JT gravity \cite{Teitelboim:1983ux}-\cite{Maldacena:2016upp} augmented by a linear deformation term $\overline{\lambda}$, where the subscript ``$\text{dJT}$'' denotes the deformed JT gravity sector. Furthermore, the boundary terms are given below
\begin{align}
S_{\text{GHY}}=-
\int_{\partial\mathcal M}
dt \sqrt{h}\,\overline{\Phi} K\hspace{1mm},\hspace{2mm}S_{\text{ct}}=\int_{\partial M} \sqrt{h}(\overline{\Phi}+\lambda) dt \label{jtbdyact}
\end{align}

On varying the deformed JT gravity action \eqref{jtact} with respect to the space-time metric and dilaton, we obtain the following equations of motion\footnote{In the rest of our analysis, we drop the overline on the deformed quantities for simplicity.}
\begin{align}
(\bigtriangledown_\mu\bigtriangledown_\nu - g_{\mu\nu}\square)\Phi + g_{\mu\nu}(\Phi + \lambda) = 0,\label{jte1}\\
     R+2 = 0.\label{jte2}
\end{align}

These equations (\ref{jte1})-(\ref{jte2}) can be solved in the static gauge \eqref{gauge}, or obtained directly by expanding the black hole solutions \eqref{bhsol1}--\eqref{bhsol2} around $r=0$ and $\theta=0$ \cite{Mahapatra:2026aor}, which yields\footnote{In Appendix D, we review another class of black hole solutions, namely the Almheiri-Polchinski (AP) model, based on the work  \cite{Almheiri:2014cka}.}
\begin{align}
    F_{\text{dbh}}(r) &= r^2 + 2\lambda r - \theta^2, \label{jtbhsol1} \\
    \Phi_{\text{dbh}}(r) &= r, \label{jtbhsol2}
\end{align}
where ``$\text{dbh}$'' denotes the deformed JT gravity black hole solutions. Moreover, the black hole horizon $r_h$ is determined by solving $F_{\text{dbh}}(r_h) = 0$, which gives
\begin{align}
    r_h = \sqrt{\lambda^2 + \theta^2} - \lambda. \label{jt_horizon}
\end{align}
It is trivial to check that in the vanishing deformation limit ($\lambda \to 0$), we recover the standard result for sine dilaton gravity \cite{Blommaert:2024ydx}.

\subsection{Thermodynamics}
Having obtained the black hole solutions for deformed JT gravity \eqref{jtbhsol1}--\eqref{jtbhsol2}, we now investigate their thermodynamic properties. In particular, we compute the Hawking temperature, Wald entropy, free energy, and examine the thermodynamic relations among them.

The Hawking temperature for deformed JT black holes is obtained by taking the derivative of $F_{\text{dbh}}(r)$ with respect to the radial coordinate \cite{Gibbons:1976ue}, which yields
\begin{align}
    T_H = \frac{1}{4\pi} F_{\text{dbh}}'(r) \Big|_{r=r_h} = \frac{\sqrt{\lambda^2 + \theta^2}}{2\pi}.\label{jthawt}
\end{align}

Now, following the approach developed for the rsDG theory in Section~\ref{sec_thermo_rsDG}, we compute the thermal observables associated with the deformed JT gravity black holes \eqref{jtbhsol1}--\eqref{jtbhsol2}. Before proceeding further, we express the radial coordinate in terms of the Hawking temperature using the expressions \eqref{jt_horizon}-\eqref{jthawt}, which yields
\begin{align}
    r_h(T_H)= 2 \pi T_H- \lambda \label{jtrh}
\end{align}

Using the definition of Wald entropy (\ref{wald}) ($\mathcal{S}_{\rm dbh}$) in the low energy regime, we obtain the following expression\footnote{Here, we drop the subscript ``$H$" for brevity.}
\begin{align}
    \mathcal{S}_{\text{dbh}} = 2\pi \Phi_{\text{dbh}}(r_h) = 2\pi\left(2\pi T-\lambda\right).\label{djt_entropy} 
\end{align}
where we use the dilaton value (\ref{jtbhsol2}) and the relation (\ref{jtrh}).

Next, we compute the free energy associated with the dJT black hole solutions. To this end, we substitute (\ref{jtbhsol1})-(\ref{jtbhsol2}) into the bulk action (\ref{jtact}) and integrate over the time and the radial coordinate, which yields
\begin{align}
    S_{\text{dJT}} = -\beta\lambda (r_c - r_h),\label{jt_bulk}
\end{align}
where $r_c$ represents the radial boundary cut-off. Notice that in the strict JT gravity limit, the physical boundary is located at infinity ($r_c \to \infty$). 

Similarly, the on-shell boundary contributions (\ref{jtbdyact}) are given by
\begin{align}
    S_{\text{GHY}} = -\beta r_c (r_c + \lambda)\hspace{1mm},\hspace{2mm} S_{\text{ct}} = \beta \left( r_c^2 + 2\lambda r_c + \lambda^2 - \frac{\lambda^2 + \theta^2}{2}\right).\label{jt_ct}
\end{align}

Adding the bulk (\ref{jt_bulk}) and boundary (\ref{jt_ct}) contributions yields the renormalized on-shell ($S_{\text{on-shell}}$) action as follows
\begin{align}
    S_{\text{on-shell}} = -\beta\left( \lambda r_h + \frac{\lambda^2 - \theta^2}{2}\right). \label{jt_ren}
\end{align}

Finally, substituting equation (\ref{jt_ren}) along with the relation (\ref{jtrh}) into the definition of renormalized free energy (\ref{free_energy_def}) for deformed JT gravity black holes gives    
\begin{align}
    \mathcal{F}_{\text{dbh}} = -\left[ 2\pi^2 T^2 - 2\pi\lambda T + \frac{\lambda^2}{2}\right] \label{djt_free_energy},
\end{align}
where we eliminate $\th$ using equation (\ref{jthawt}).

It is interesting to note that the Wald entropy \eqref{djt_entropy} and free energy \eqref{djt_free_energy} satisfy the standard thermodynamic relation
\begin{align}
    \frac{d \mathcal{F}_{\text{dbh}}(T)}{dT} = -\mathcal{S}_{\text{dbh}}(T).
\end{align}
\begin{figure}[htp]
    \centering
    \begin{subfigure}[b]{0.48\textwidth}
        \centering
        \includegraphics[width=\textwidth]{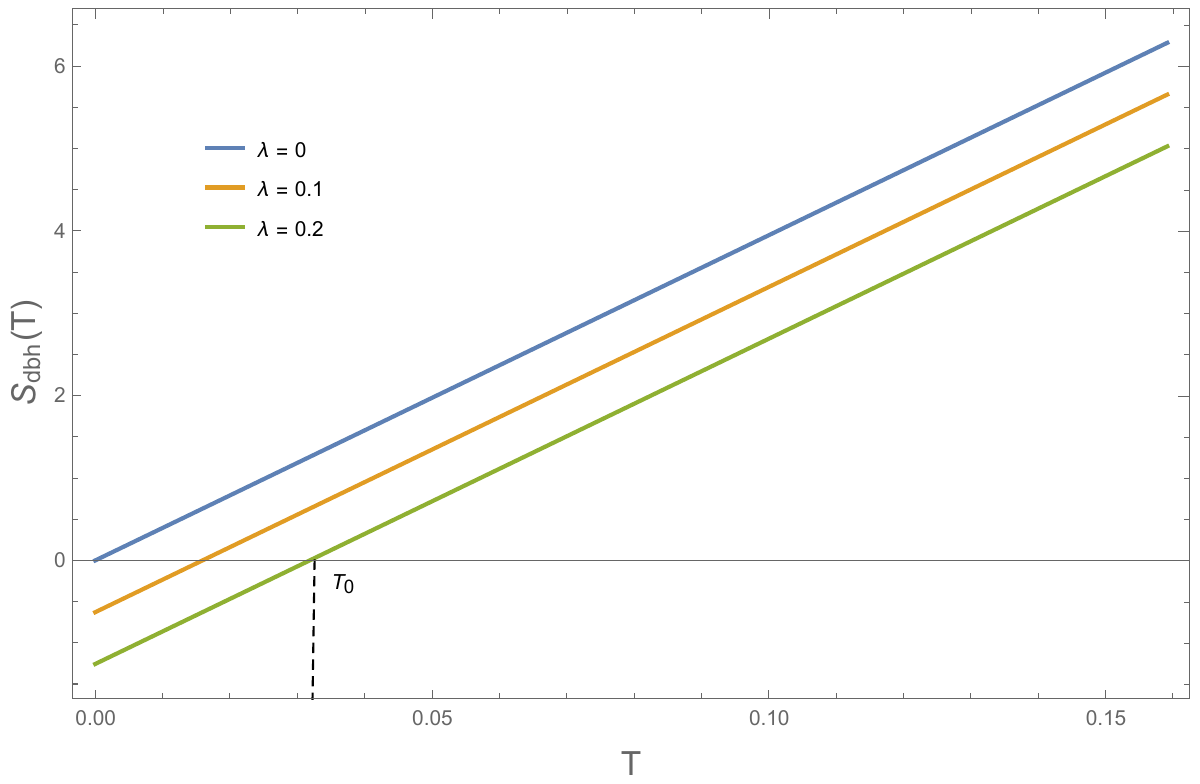}
        \caption{Entropy vs temperature}
        \label{JTEimg}
    \end{subfigure}
    \hfill
    \begin{subfigure}[b]{0.48\textwidth}
        \centering
        \includegraphics[width=\textwidth]{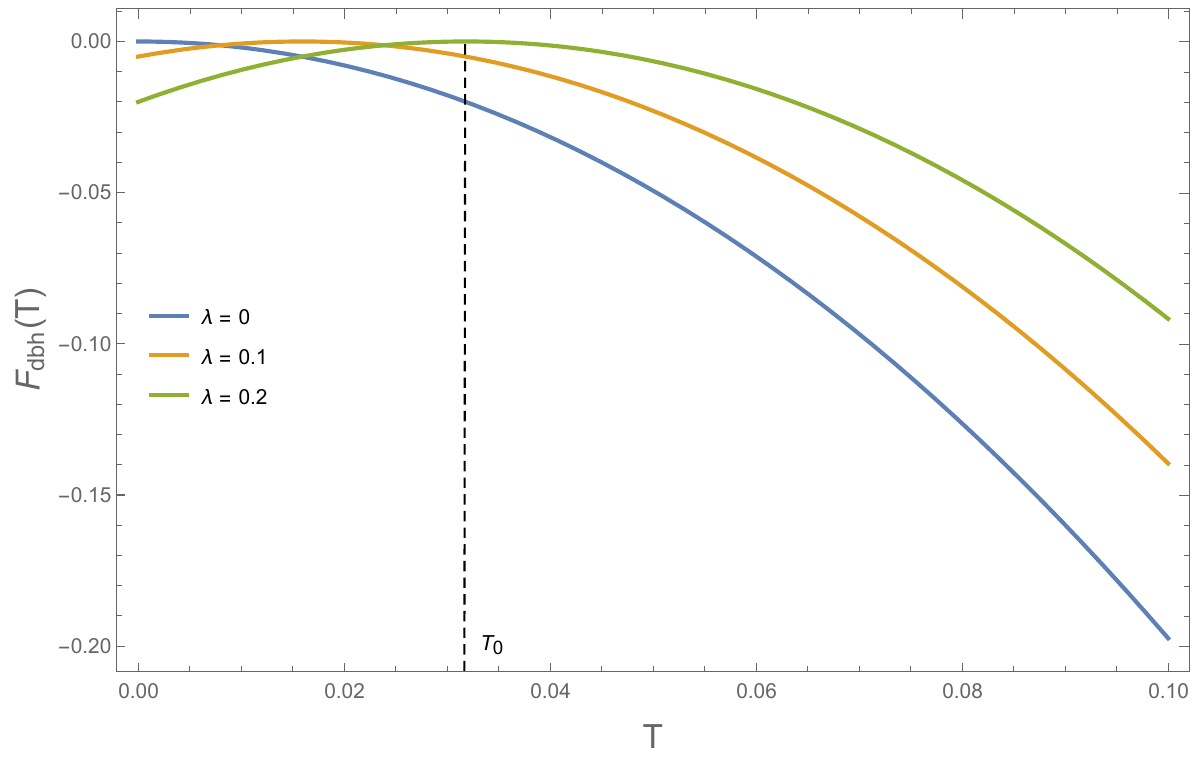}
        \caption{Free energy vs temperature}
        \label{JTFimg}
    \end{subfigure}
    \caption{Phase stability of dJT black holes for various deformation parameter.}
    \label{JTfig:entropy_free_energy}
\end{figure}

Furthermore, the expressions \eqref{djt_entropy} and \eqref{djt_free_energy} indicate the existence of a critical temperature, $T_0 = \frac{\lambda}{2\pi}$, at which the Wald entropy vanishes. In the low-temperature regime ($T < T_0$), the system enters an unphysical phase, where the entropy becomes negative. At the same time, the free energy increases with rise in temperature, reaching a local maximum at the threshold, $T=T_0$ as illustrated in Figure \ref{JTfig:entropy_free_energy}.

As previously discussed in Section \ref{sec_thermo_rsDG}, this low temperature domain ($T < T_0$) corresponds to an unphysical number of degrees of freedom ($\Omega(T)$) as shown below
\begin{align}
    \Omega(T)=\text{exp}(\mathcal{S}_{\text{dbh}})=\text{exp}[2\pi(2\pi T-\lambda)], \label{jtdensity}
\end{align}
where the Wald entropy $(\mathcal{S}_{\text{dbh}}) $ for deformed JT black hole solutions is provided in equation (\ref{djt_entropy}). Expanding $\Omega(T)$ around the critical temperature, i.e. $T=T_0-\epsilon$, with $0<\epsilon<<1$ is a small parameter, yields
\begin{align}
    \Omega(T_0-\epsilon)=1-4\pi^2\epsilon+\mathcal{O}(\epsilon^2),
\end{align}
which remains strictly less than unity, confirming the presence of unphysical states.

Conversely, for $T > T_0$, the Wald entropy \eqref{djt_entropy} grows linearly with temperature, while the free energy \eqref{djt_free_energy} decreases monotonically that ensures a local thermally stable phase (see Figure \ref{JTfig:entropy_free_energy}). In this high-temperature regime ($T > T_0$), the thermodynamic observables of dJT black holes exhibit either linear \eqref{djt_entropy} or quadratic \eqref{djt_free_energy} dependence. In contrast, rsDG black holes display a significant non-linear temperature dependence across both the entropy \eqref{ET} and free energy \eqref{FT}, as demonstrated in Figure \ref{fig:entropy_free_energy}. For instance, compared to dJT black holes, the entropy of rsDG black holes increases more rapidly, while the corresponding free energy decreases more gradually with rising temperature. These non-linear corrections reflect the fact that rsDG theory provides an ultraviolet (UV) completion and possesses a richer phase structure.

\section{Conclusion and future directions}

In the present work, we have explored the phase stability of sine dilaton gravity under marginal deformations. We began by reviewing the background within the context of Liouville theory and subsequently introduced these deformations into sDG following the approach of \cite{Giribet:2024men}. In this framework, we derived the black hole solutions associated with the deformed sDG and examined their thermodynamic phase stability. 

Our analysis reveals that the marginal deformation significantly alters the thermodynamic observables, notably introducing a threshold minimum temperature $(T_0)$. Below this critical temperature ($T < T_0$), the system encounters an unphysical phase where the Wald entropy becomes negative and the density of states drops below unity, thereby reducing the number of accessible physical states. Conversely, above the critical temperature ($T > T_0$), the system exhibits a stable black hole phase. 

Furthermore, we computed the Casimir function for the deformed sDG and identified it with the ADM mass. Remarkably, we found that the temperature, entropy, and free energy are modified in such a way that the first law of thermodynamics remains invariant compared to the undeformed theory. Finally, we performed a similar analysis in the IR limit of the deformed theory, which yields standard JT gravity accompanied by the deformation. The black hole solutions obtained in this deformed JT limit and their subsequent thermal properties confirm that the deformation continues to induce a minimum temperature threshold below which all thermodynamic observables become physically inadmissible.

Below we provide a few interesting directions for future extensions.
\begin{itemize}
    \item In this work, our analysis is performed entirely in the bulk, where the marginal deformation introduces a minimum temperature cut-off that modifies the complete thermodynamics of the system. Within the framework of the gauge/gravity duality \cite{Maldacena:1997re}, it would be an interesting direction to investigate similar behavior in the dual one-dimensional deformed DSSYK model. In particular, this deformation is expected to modify the chord Hilbert space by introducing a cut-off, thereby reducing the number of physically admissible states.
    \item It would also be worthwhile to study the implications of the deformation on various holographic observables, such as the two-point correlation function \cite{Goulian:1990qr}
, the boundary stress-energy tensor, and the central charge \cite{Hartman:2008dq}-\cite{Rathi:2023vhw}.
\end{itemize}

\section*{Acknowledgments}
    PM and DR are indebted to the authorities of Indian Institute of Technology, Roorkee for their unconditional support towards researches in basic sciences. HR would like to thank the authorities of K.R. Mangalam University, Gurugram, for their support. DR also acknowledges the Mathematical Research Impact Centric Support (MATRICS) grant (MTR/2023/000005) received from ANRF, India.

\appendix
\section{Single marginal deformation}
In this section, we consider the case of a single marginal deformation \cite{Giribet:2024men} in the context of sine dilaton gravity. In particular, we compute the black hole solutions associated with the complex deformation and examine their thermodynamic properties, such as the Hawking temperature, Wald entropy and the free energy.

We begin with the following action \cite{Giribet:2024men}
\begin{equation}
    S = -\frac{1}{4|\log q|}\int_M d^2x\,\sqrt{g}\,
    \left[
        \Phi R
        +2\sin\Phi
        +\lambda e^{i\Phi}
    \right]
    +S_{\rm GHY}+S_{\rm ct},
    \label{full_deformact}
\end{equation}
where the boundary terms are given below
\begin{equation}
S_{\rm GHY}
=
-\frac{1}{2|\log q|}
\int_{\partial\mathcal M}
dt \sqrt{h}\,\Phi K\hspace{1mm},\hspace{2mm}S_{ct}=\frac{i}{2|\log q|}
\int_{\partial M}
d\tau\,
\sqrt{h}\,
e^{-i\Phi/2}. \label{ccounter}
\end{equation}

After varying the action \eqref{full_deformact}, we obtain the following equations of motion
\begin{align}
\nabla_\mu\nabla_\nu\Phi-g_{\mu\nu}\Box\Phi-\frac12 g_{\mu\nu}\mathcal{V}(\Phi)=0,\label{eq_f1}\\
R - \mathcal{V}'(\Phi)=0,\label{eq_f2}
\end{align}
where we define the potential as $\mathcal{V}(\Phi) =  -2\sin\Phi-2\lambda e^{i\Phi}$. One can simplify the above equations (\ref{eq_f1})-(\ref{eq_f2}) in the Euclidean static gauge (\ref{gauge}) to find the following black hole solutions
\begin{align}
F(r) &= 2\cos\theta
    -2\cos r
    -i\lambda e^{ir}, \label{f_bhsol1} \\
\Phi(r) &= r. \label{f_bhsol2}
\end{align}

The black hole horizon ($r_h$) can be obtained by solving $ F(r_h)=0$, yielding
\begin{align}
    r_h = \cos^{-1}\left( \frac{\cos\theta}{\sqrt{1+ i\lambda}}\right)+\frac{i}{2}\log(1+i\lambda). \label{rhc}
\end{align}
It is trivial to check that in the limit $\lambda \rightarrow 0$, we recover the undeformed horizon, $r_h = \theta$ \cite{Blommaert:2024ydx}. 

Furthermore, using a suitable coordinate transformation \cite{Blommaert:2024ydx},
\begin{align}
     r = \frac{\pi}{2} + i\log \left( \rho+i\cos\theta \right),
    \label{eq:coordinate_transformation}
\end{align}
the space-time metric (\ref{gauge}) can be recast in a conformally $\text{AdS}_2$ form as
\begin{align}
    ds^2 = e^{i r(\rho)} \left[ (\rho^2 - \rho_h^2) d\tau^2 + \frac{d\rho^2}{\rho^2 - \rho_h^2} \right],
\end{align}
where $\rho_h = \left(\sin^2 \theta + i\lambda \right)^{1/2}$ denotes the horizon radius in $\rho$ coordinate. It is worth noting that although the space-time metric is conformally $\text{AdS}_2$, it remains a complex quantity due to the complex nature of the horizon radius ($\rho_h$). Consequently, the holographic dictionary corresponding to the boundary dDSSYK theory becomes non-trivial.

 Next, using (\ref{f_bhsol1}), we compute the Hawking temperature ($T$) for the complex deformed black holes as follows
 \begin{align}
     T = \frac{F'(r_h)}{4\pi} =  \frac{2\sin r_h + \lambda e^{ir_h}}{4\pi}.\label{tempapp}
  \end{align}
  
One can also express the Hawking temperature ($T$) given in (\ref{tempapp}) in terms of $\theta$ and $\lambda$, as discussed earlier in (\ref{hwtemp2}). Towards this end, we rewrite $F'(r_h)$ as
\begin{align}
    F'(r_h)= -i(1+i\lambda)e^{ir_h} + ie^{-ir_h}, \label{fprime}
\end{align}
where we have used the trigonometric identity $\sin r_h = (e^{ir_h}-e^{-ir_h})/2i$. Furthermore, using the horizon condition $F(r_h) = 0$ (\ref{f_bhsol1}) along with the identity $\cos r_h = (e^{ir_h} + ie^{-ir_h})/2$, we find  \begin{align}
      (1+i\lambda)e^{ir_h} =   2\cos \theta - e^{-ir_h}. \label{horizon_condition1}
  \end{align}
  
Now, substituting (\ref{horizon_condition1}) into (\ref{fprime}) simplifies the expression to
\begin{align}
    F'(r_h)= 2i(e^{-r_h} - \cos \theta). \label{fprime2}
\end{align}
Next, taking the square of equation (\ref{fprime2}) and using the expression (\ref{horizon_condition1}), one obtains
\begin{align}
    F'(r_h)^2 = 4(\sin^2 \theta + i\lambda). \label{fprime3}
\end{align}
Finally, plugging (\ref{fprime3}) into (\ref{tempapp}), we arrive at the following expression for Hawking temperature
\begin{align}
    T = \frac{\sqrt{\sin^2 \theta + i\lambda}}{2\pi}. \label{hwtempc}
\end{align}
  
Next, we study the thermal properties associated with the complex deformed black holes. Before proceeding further, we first express the horizon radius (\ref{rhc}) in terms of Hawking temperature (\ref{hwtempc}). To begin with, we rewrite (\ref{hwtempc}) as follows
\begin{align}
    \cos \theta = \sqrt{1-4\pi^2T^2 + i\lambda}, \label{thetaexp2}
\end{align}
where we have used the trigonometric identity $\sin^2\th+\cos^2\th=1$.

Now, substituting (\ref{thetaexp2}) in (\ref{rhc}), the horizon radius, $r_h(T)$ can be expressed as
\begin{align}
    r_h(T)= \cos^{-1}\left( \frac{\sqrt{1-4\pi^2T^2 + i\lambda}}{\sqrt{1+i\lambda}}\right) + \frac{i}{2} \log(1+i\lambda). \label{rhc2}
\end{align}

Having obtained the black hole solutions (\ref{f_bhsol1})-(\ref{f_bhsol2}) and the horizon radius (\ref{rhc2}) for complex deformed sine-dilaton gravity, we now aim to study their thermodynamics. To begin with, we first examine the Wald entropy (\ref{wald}) $(S_{\text{complex}}(T))$, which can be obtained by varying the action (\ref{full_deformact}) with respect to the Riemann tensor. This yields
\begin{align}
    S_{\text{complex}}(T)=\frac{\pi \Phi(r_h(T))}{|\log q|} =&\hspace{1mm} \frac{\pi}{|\log q|}
\Bigg[
\cos^{-1}\left(
\frac{\sqrt{1-4\pi^2T^2+i\lambda}}
{\sqrt{1+i\lambda}}
\right)+\frac{i}{2}\log(1+i\lambda)
\Bigg].\label{coment}
\end{align}

Next, we compute the free energy ($\mathcal{F}_{\text{complex}}(T)$) for the complex deformed black holes. To this end, we substitute the solutions (\ref{f_bhsol1})-(\ref{f_bhsol2}) into (\ref{full_deformact}) and evaluate the on-shell bulk action ($S^{\text{on-shell}}$) after integrating over the radial and time coordinates, which yields
\begin{align}
S^{\text{on-shell}} = -\frac{\beta}{4|\log q|}\left[B(r_c)-B(r_h),\label{cbulk} \right]
\end{align}
where we define $B(r) = -2r\sin r - 2\cos r - \lambda r e^{ir} - 2i\lambda e^{ir}$. Here $r_c$ denotes the boundary cut-off. Similarly, we obtain the boundary contribution, which is given below
\begin{align}
    S_{\text{GHY}}^{\text{on-shell}} = -\frac{\beta}{4|\log q|}\left[r_c F'(r_c) \right]\hspace{1mm},\hspace{2mm}
    S_{\text{ct}}^{\text{on-shell}}= - \frac{\beta}{2|\log q|}\left[ r_h\left( \frac{2\sin r_h + \lambda e^{i r_h}}{2}\right) + 2\cos r_h 
    + i\lambda e^{i r_h} \right]. \label{renc}
\end{align}

Finally, by adding equations (\ref{cbulk}) and (\ref{renc}), and substituting the horizon radius expression from equation (\ref{rhc2}) into the free energy definition (\ref{free_energy_def}), we obtain
\begin{align}
    \mathcal{F}_{\text{complex}}(T) =& \hspace{1mm} - \frac{1}{2|\log q|}
\Bigg(
2\pi T
\left[
\cos^{-1}\left(
\frac{\sqrt{1-4\pi^2T^2+i\lambda}}
{\sqrt{1+i\lambda}}
\right)
+\frac{i}{2}\log(1+i\lambda)
\right]+2\sqrt{1-4\pi^2T^2+i\lambda}
\Bigg).\label{compfreeapp}
\end{align}

Notice that the Hawking temperature \eqref{hwtempc}, the Wald entropy \eqref{coment}, and the free energy \eqref{compfreeapp} are all complex quantities for the deformation under consideration \eqref{full_deformact}, which renders them physically inadmissible. However, it is interesting to notice that if we rescale the deformation parameter, i.e., $\lambda\rightarrow i\lambda$ and consider only the real part, the Hawking temperature \eqref{hwtempc} and the Wald entropy \eqref{coment} become
\begin{align}
    T = \frac{\sqrt{\sin^2 \theta-\lambda}}{2\pi}\hspace{1mm},\hspace{2mm} 
    S = \frac{\pi}{|\log q|}\cos^{-1}\left( \frac{\cos \theta}{\sqrt{1-\lambda}}\right). \label{coment2}
\end{align}

Notice that the ADM mass of the theory remains unchanged (i.e, $M = -\frac{\cos \theta}{2|\log q|}$) as discussed in appendix C. One can easily check that equation (\ref{coment2}) satisfy the first law of black hole mechanics, i.e, $TdS = dM$ holds identically.
Motivated by this observation, to bypass complex thermodynamic observables, in Section \ref{Section real} we consider a marginal deformation defined as the sum of the deformation in \eqref{full_deformact} and its complex conjugate, thereby yielding real deformations. As discussed in Section \ref{sec_thermo_rsDG}, this real deformation produces entirely real thermodynamical observables.  

\section{Geometric identity}
In this Appendix, we derive the geometric relation $X^+ X^- = -\frac{1}{2} (\nabla X)^2$, which plays a crucial role in constructing the Casimir invariant for real deformed sine dilaton gravity (rsDG).

In the Grumiller Kummer Vassilevich (GKV) formalism \cite{Grumiller:2002nm}, the target space light-cone coordinates $X^a$ ($a \in \{+, -\}$) can be expressed in differential form as
\begin{align}
    X^a = *\left( e^a \wedge dX \right),
    \label{eq:Xa_identity}
\end{align}
where $*\,$ denotes the two dimensional Hodge dual operator, $e^a = e^a_\mu dx^\mu$ represents the frame two-bein (zweibein) 1-forms, and $dX = \partial_\mu X dx^\mu$ is the exterior derivative of the scalar dilaton field $X$.

In component form, the above expression (\ref{eq:Xa_identity}) simplifies \cite{Grumiller:2002nm} to
\begin{equation}
    X^a = \frac{\tilde{\epsilon}^{\mu\nu}}{e} e^a{}_\mu \partial_\nu X,\label{compcord}
\end{equation}
where $\tilde{\epsilon}^{\mu\nu}$ is the two dimensional Levi-Civita symbol\footnote{ Here, we use the notation, $\tilde{\epsilon}^{01} = 1.$}, and $e = \det(e^a{}_\mu)$ is the determinant of the zweibein frame.

Now, we define the norm of the coordinates (\ref{compcord}) as follows
\begin{align}
X^aX_a
&=
\eta_{ab}
X^aX^b
\nonumber\\
&=
\eta_{ab}
\frac{\tilde{\epsilon}^{\mu\nu}}{e}
\frac{\tilde{\epsilon}^{\rho\sigma}}{e}
e^a{}_\mu
e^b{}_\rho
\partial_\nu X
\partial_\sigma X
\nonumber\\
&=
\frac{\tilde{\epsilon}^{\mu\nu}
\tilde{\epsilon}^{\rho\sigma}}
{e^2}
g_{\mu\rho}
\partial_\nu X
\partial_\sigma X .
\end{align}
Using the two-dimensional identity \cite{nakahara2003geometry}
\begin{equation}
    \frac{\tilde{\epsilon}^{\mu\nu} \tilde{\epsilon}^{\rho\sigma}}{e^2} g_{\mu\rho} = -g^{\nu\sigma},
\end{equation}
one finds
\begin{equation}
    X^a X_a = -(\nabla X)^2.
    \label{XaXa}
\end{equation}
Finally, substituting the relation $X^a X_a = 2 X^+ X^-$ \cite{Grumiller:2002nm} into (\ref{XaXa}) yields
\begin{equation}
    X^+ X^- = -\frac{1}{2} (\nabla X)^2.
    \label{XpXm}
\end{equation}

\section{Exterior derivative of the Casimir function}
\label{app:exterior_derivative}

In this Appendix, we explicitly verify that the exterior derivative of the Casimir function ($C$) \eqref{casimir_rsDG} for real marginally deformed sine dilaton gravity vanishes on-shell i.e., $dC = 0$. 

The exterior derivative of the Casimir invariant is given by
\begin{equation}
dC = \partial_{\alpha} C \, dX^{\alpha}.
\end{equation}
Differentiating $C$ \eqref{casimir_rsDG} with respect to space-time coordinates ($X^{\alpha}$) yields
\begin{equation}
\partial_{\alpha} C = -(\nabla_{\alpha} \nabla_{\mu} X) \nabla^{\mu} X + (\sin X + \lambda \cos X) \nabla_{\alpha} X.
\label{eq:dC_component_app}
\end{equation}

To evaluate the above expression \eqref{eq:dC_component_app} on-shell, we take the trace of the metric equation of motion \eqref{eq1}, which gives
\begin{equation}
\Box X = 2(\sin X + \lambda \cos X),
\label{eq:trace_eom_app}
\end{equation}
where we use the two dimensional identity, i.e., $g^{\mu\nu} g_{\mu\nu} = 2$.

Substituting the trace condition \eqref{eq:trace_eom_app} back into the equation of motion \eqref{eq1} reduces it to the following
\begin{equation}
\nabla_{\alpha} \nabla_{\mu} X = g_{\alpha \mu} (\sin X + \lambda \cos X).
\label{eq:eom_identity_app}
\end{equation}

Finally, substituting the above relation (\ref{eq:eom_identity_app}) into the equation \eqref{eq:dC_component_app}, we find exact cancellation between the kinetic and potential contributions as shown below
\begin{equation}
\partial_{\alpha} C = -(\sin X + \lambda \cos X) \nabla_{\alpha} X + (\sin X + \lambda \cos X) \nabla_{\alpha} X = 0.
\end{equation}

Consequently, $dC = 0$ holds throughout space-time, implying conservation of the ADM mass given in equation (\ref{mass_rsDG}). This conservation law is independent of the real marginal deformation parameter $\lambda$.

\section{Deformed AP model}
In this section, we present a distinct class of solutions associated with deformed JT gravity \eqref{jtact}. In particular, we obtain solutions analogous to the Almheiri-Polchinski black holes in two spacetime dimensions \cite{Almheiri:2014cka}. However, their thermodynamic behavior is identical to the results presented in Section \ref{sectiondjt}, as given in equations \eqref{jtbhsol1}-\eqref{jtbhsol2}.

For completeness, we rewrite the equations of motion for dJT gravity \eqref{jte1}-\eqref{jte2} as
\begin{align}
(\bigtriangledown_\mu\bigtriangledown_\nu - g_{\mu\nu}\square)\Phi + g_{\mu\nu}(\Phi + \lambda) = 0,\label{apjte1}\\
     R+2 = 0.\label{apjte2}
\end{align}

To solve the equations of motion \eqref{apjte1}--\eqref{apjte2}, we adopt the static Euclidean ansatz \cite{Almheiri:2014cka}
\begin{align}
ds_E^2 = e^{2\omega(z)} \left( d\tau^2 + dz^2 \right), \label{apgauge}
\end{align}
where $\tau$ is the Euclidean time, $z$ is the radial coordinate, and $\omega(z)$ is the conformal factor. In this gauge, the explicit solutions are given by
\begin{align}
e^{\omega(z)} = \frac{\rho_h}{\sinh(\rho_h z)}, \quad \Phi(z) = \rho_h \coth(\rho_h z) - \lambda,\label{apsola}
\end{align}
where $\rho_h$ denotes the horizon location and the boundary is located at $z=0$.

Alternatively, employing the coordinate transformation \cite{Almheiri:2014cka}
\begin{align}
\rho = \rho_h \operatorname{coth}(\rho_h z)
\label{eq:coord_trans}
\end{align}
in the line element \eqref{apgauge}, we obtain the standard Euclidean Schwarzschild metric given by
\begin{align}
ds_E^2 = (\rho^2 - \rho_h^2) d\tau^2 + \frac{d\rho^2}{\rho^2 - \rho_h^2}, \quad \Phi = \rho - \lambda.\label{schsol}
\end{align}
This demonstrates that the hyperbolic Almheiri--Polchinski solution (\ref{apsola}) and the Schwarzschild solution 
(\ref{schsol}) are related by a coordinate transformation and thus describe identical physical sectors, as further explored in Section \ref{sectiondjt}.

\end{document}